\documentclass[draft
    ,final            % use final for the camera ready runs
  ]
  {aipproc}

\layoutstyle{6x9}

\begin{document}

\def\plotfiddle#1#2#3#4#5#6#7{\centering \leavevmode
\vbox to#2{\rule{0pt}{#2}}
\includegraphics{#1}}
%
%    \plotfiddle{EPSFILE}{VSIZE}{ROT}{HSF}{VSF}{HTRANS}{VTRANS}
%
%    VSIZE      vertical white space to allow for plot
%    ROT        rotation angle
%    HSF        horiz scale factor
%    VSF        vert scale factor
%    HTRANS     horiz translation
%    VTRANS     vert translation

\newcommand{\HI}{\mbox{H\,{\sc i}}}
\newcommand{\HII}{\mbox{H\,{\sc ii}}}

\title{Growth and Destruction of Disks: \\ Combined \HI\ and \HII\ View}

\classification{}

\keywords      {}

\author{Matthew Bershady}{
  address={Astronomy Department, University of Wisconsin-Madison}
}
\author{Marc Verheijen}{
  address={Kapteyn Astronomical Institute, University of Groningen}
}
\author{Steven Crawford}{
  address={South African Astronomical Observatory, Cape Town}
}

\begin{abstract}

  How large disk galaxies have evolved in, and out of, the blue cloud
  of actively star-forming galaxies as a function of environment and
  time is an outstanding question. Some of the largest disks become
  systems like M31, M33 and the Milky Way today. In denser
  environments, it appears they transform onto the red sequence.
  Tracking disk systems since z<0.5 as a function \HI\ mass, dynamical
  mass, and environment should be possible in the coming decade. \HI\
  and optical data combined can sample outer and inner disk dynamics
  to connect halo properties with regions of most intense
  star-formation, and the gas reservoir to the consumption rate.  We
  describe existing and future IFUs on 4-10m telescopes that
  complement upcoming \HI\ surveys for studying disks at z<0.5. Multiple
  units, deployable over large fields-of-view, and with logarithmic
  sampling will yield kinematic and star-formation maps and properties
  of the stellar populations, resolving the core but retaining
  sensitivity to disk outskirts.

\end{abstract}

\maketitle

%%%%%%%%%%%%%%%%%%%%%%%%%%%%%%%%%%%%%%%%%%%%
%% MAINMATTER
%%%%%%%%%%%%%%%%%%%%%%%%%%%%%%%%%%%%%%%%%%%%

\section{The Blue Cloud and Red Sequence}

A general but powerful statement about galaxy evolution stems from the
results of merger simulations (e.g., Mihos \& Hernquist 1994): Disks
are fragile, destroyed in major mergers, and heated in minor
mergers. It has been suggested that wet (gas-rich) mergers may
complicate this picture, in that wet mergers can make new disks
(Hammer et al. 2005, Springel \& Hernquist 2005). To yield old stellar
populations in dynamically-cold, star-forming disks today such
wet-merging must happen early. We can therefore conclude that over the
last Hubble time large disks we see today have been stirred, not
shaken; they have tidal streams but are not train-wrecks.

However, some disk galaxies do evolve from the blue cloud to form the
red sequence. Indeed, most of the stars made in the blue-cloud
population end up in the red sequence (e.g., Bell et al. 2007). Much of
this transformation appears to hinge on local environment. A tractable
problem is to identify which galaxies have been stirred and which
shaken (or stripped) at each cosmic epoch, and tag their environment.
Some of this tagging can be done purely from optical photometrics, as
we illustrate in Figure 1. For those only stirred, we can then monitor
statistically their smooth and continuous accretion (growth and gas
supply) and star-formation (gas consumption). To make this last
critical step requires connecting optical and \HI\ views, and ultimately
also tying in a picture of the molecular gas distribution.

Considerable attention in the SKA science-literature is paid to
connecting the \HI\ mass-function evolution and the comoving
star-formation rate (e.g., van der Hulst et al. 2004). At this
conference, Zwaan has reported a clear picture on the \HI\ mass function
and its evolution, or lack thereof. However, do we know if these
trends of mass (fuel) and star-formation (consumption) are
self-consistent? To get a handle on this, we need to unpack the
comoving quantities (e.g., Heavens et al. 2004), and look specifically
at the multivariate distributions over epoch, such as the
star-formation rates (SFR) as function of both dynamical and \HI\ mass.
Likewise we need to identify the individual galaxies, e.g., at the
knee of the \HI\ mass-function, which we can presume are likely to
include many of the large-disk systems.

\begin{figure}
\plotfiddle{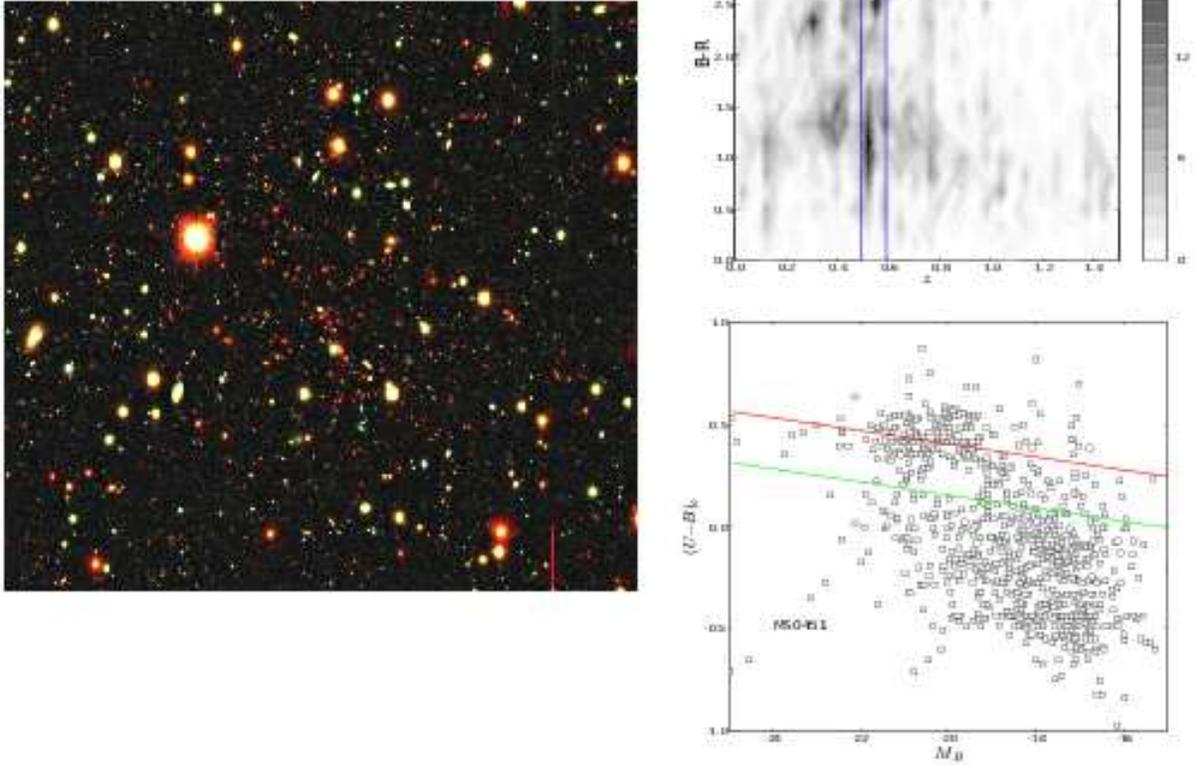}{4in}{0}{90}{90}{-500}{-180}
\caption{The blue cloud and red sequence in a deep field around
  MS0451, a rich cluster ($\sigma = $ 1354 km s$^{-1}$ and L$_x$ =
  40e44 ergs s$^{-1}$) at z=0.54 from the the WIYN Long-Term
  Variability Survey ($UBRIz$ bands, Crawford et al. 2006, 2008)
  The $UBI$ 3-color image at left is a 10x10 arcmin field. The
  corresponding $B-R$ color--photometric-redshift diagram illustrates
  isolation of the cluster volume and the surrounding environment. The
  rest-frame $U-B$ color-magnitude diagram shows the blue cloud and red
  sequence in this redshift slice. Data like these are at the
  upper-redshift limit of what could be well-probed with deep SKA-precursor
  \HI\ surveys.}
\end{figure}

These broad brush-strokes motivate a program to trace a set of key
observables in the blue cloud across comic time. The observables
include dynamical mass, baryonic mass (\HI\, molecular, and stellar),
SFR, and abundances. These allow us to frame a model of 
baryon processing in gravitational wells, and constrain the efficiency
of this processing. At our disposal are the now well-known
correlations between dynamical mass, baryonic mass (light), and
metallicity. Let us assume that molecular gas is estimated from the
atomic, or measured with ALMA, and that stellar mass is dynamically
calibrated via low-redshift studies such as the DiskMass survey
(Verheijen et al. 2007a). Here we focus on the optical instrumentation
needed to connect future \HI\ surveys with spatially resolved
star-formation rates and gas-phase abundances.

\section{\HI\ View: State of the Art}

Deep aperture-synthesis surveys with Westerbork and the VLA have now
mapped \HI\ to z=0.2 at masses well below M* (Verheijen et
al. 2007b). The extensive scientific dividends from such studies are
discussed in these proceedings by van Gorkom. These impressive surveys
have solid detections for 42 sources in 2 x 0.4 deg$^2$ fields, and
expect 200 sources in 1000 hours of integration.

% Give depth of \HI\ observation

With dedicated, deep surveys using SKA path-finders (e.g., eVLA,
MeerKAT, ASKAP) we can expect to extend to $z\leq0.5$. The meat of
these surveys (where the \HI\ detections go well below M*) will be in
something like the Sloan Volume, i.e., with a characteristic redshift
of 0.1-0.2.  Optical follow-up, especially for emission-line science
(kinematics, abundances, SFR) are well in the domain of 4m-class
telescopes.

As an example of the power in optical follow-up, Figure 2 contrasts
state-of-the-art \HI\ and H$\alpha$ kinematic data for a cluster galaxy
at z=0.2. The optical data offers a spatially detailed supplement
which connects the \HI\ map of the fuel reservoir and mass enclosed on
large scales to the dynamical heart of the galaxy where the fuel is
being processed. This is achieved in the optical in a very modest
amount of time per target.

\begin{figure}
\plotfiddle{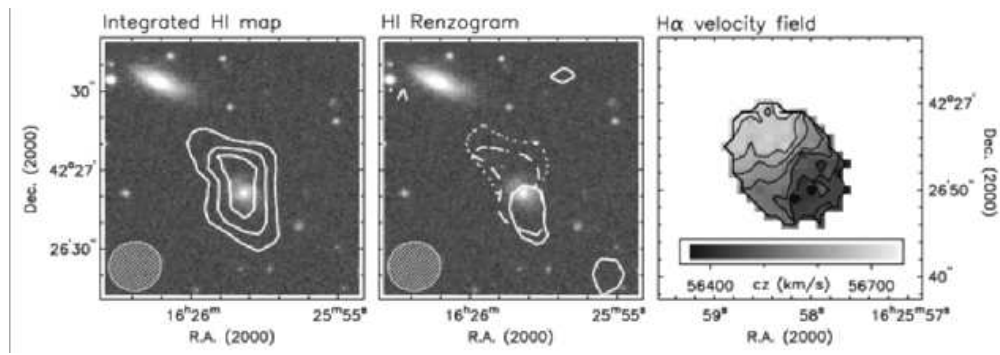}{1.5in}{0}{70}{70}{-425}{-200}
\caption{\HI\ and H$\alpha$ internal-velocity structure in a cluster
  galaxy at z=0.2 in A2192 (Verheijen 2004). \HI\ data is typical of
  what can be achieved with today's deepest surveys in advance of
  SKA-pathfinder facilities such as eVLA, ASAKP and MeerKAT. The
  H$\alpha$ velocity-field was taken in 80 minutes on the Calar Alto
  3.5m telescope using a 16x16 1'' element integral field-unit of
  PMAS.}
\end{figure}

\section{Matching the State of the Art in the Optical}

The challenge for matching \HI\ surveys with optical spectroscopy is
field-of-view.  While the sensitivity per target (or spaxel) is
substantially higher in the optical, radio aperture-synthesis
telescopes have significantly larger primary beams than most optical
spectrographs. Since the science signal is only coming from a small
portion of the solid angle, what is needed most importantly are
optical spectrographs with large patrol fields. This has been
achieved, traditionally, with fibers. A number of telescopes have such
capability. Here we focus on the WIYN 3.5m telescope, the multi-fiber
positioner Hydra (Barden et al. 1993), and the spectrograph which this
feeds: the Bench.

% Rebuilding the WIYN Bench Spectrograph

\begin{figure}
\plotfiddle{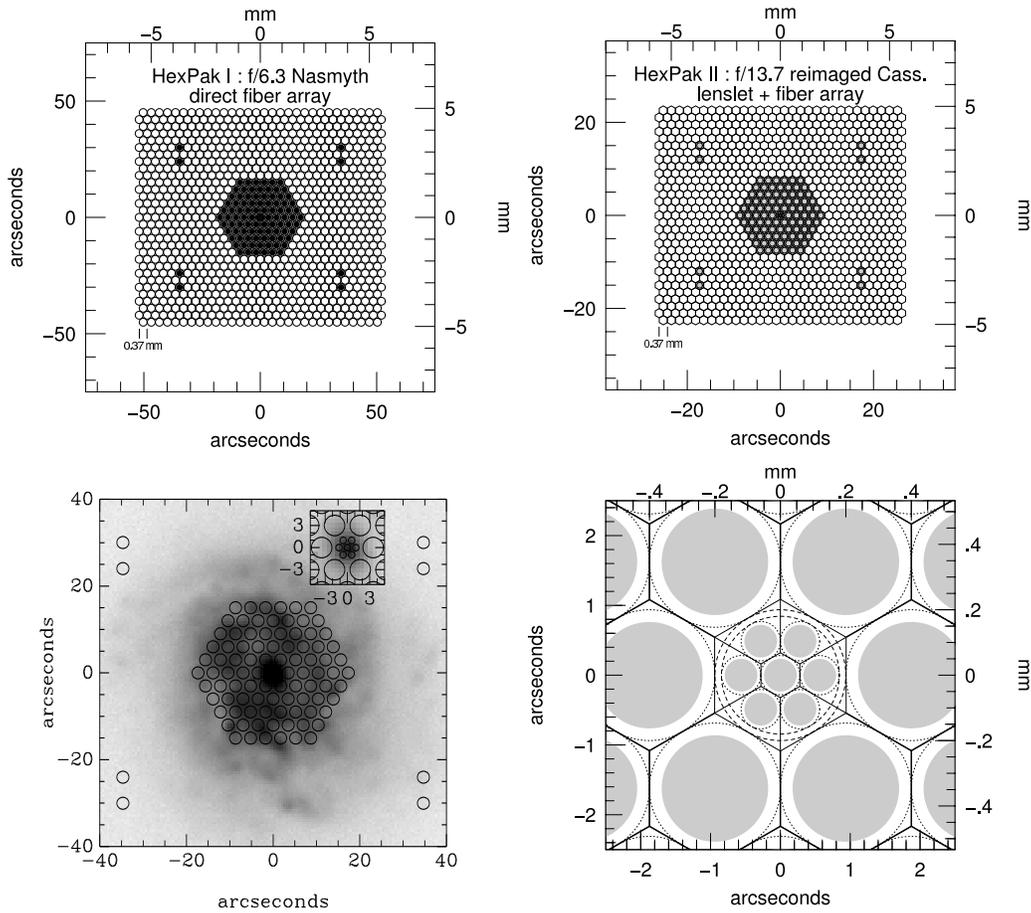}{4in}{0}{70}{70}{-430}{-80}
\caption{HexPak~I \& II, logarithmic-sampled IFUs proposed for WIYN,
  well-suited for ALFALFA survey follow-up. {\it Top Left.} HexPak~I
  bare-fiber IFU for the f/6.3 Nasmyth port, sampling at 2.''81 over
  30'' is a replacement of, and improvement on the former DensePak
  IFU.  Open fibers are for mechanical packing and buffering. {\it
    Bottom Left.} HexPak~I overlay on Seyfert galaxy NGC 3982 in Ursa
  Major, with an inset shows sampling of AGN core. {\it Top Right.}
  HexPak~II fiber+lenslet IFU for the f/13.7 reimaged-Cassegrain port,
  sampling at 1.''29 over 15''.  {\it Bottom Right.} Blow-up of inner
  2-arcsec region showing variable-pitch region of array for
  HexPak~II. Shaded regions are active fiber areas which are fed by an
  integrally-covering array of hexagonal and trapezoidal lenslets.
  Fibers are a new broad-spectrum product with transmission
  performance surpassing older fibers in both blue and red.}
\end{figure}

Hydra has a superb advantage for \HI\ survey follow-up due to its 1
degree patrol field. The problem with WIYN multi-fiber spectroscopy
has been that the Bench Spectrograph, while used for roughly 65\% of
the observing time only has 3-4\% throughput. The solution has been to
undertake a major, multi-year effort to rebuild the spectrograph.

Bench Spectrograph improvements include two new high-throughput
volume-phase holographic (VPH) gratings; a new, high-QE, low-noise
CCD; and a new, high-throughput and faster refractive collimator. With
the exception of the new collimator (due to be commissioned in
late-summer to early Fall 2008), all components are now in place. The
total gains over the original system are up to a factor 3.5 improved
efficiency with little-to-no loss in spectral resolution (due to
better sampling and improved optical image quality). This yields a
competitive total system efficiency of 10\% at $\lambda/\Delta\lambda
= 20,000$ (echelle), and 15\% at $\lambda/\Delta\lambda = 10,000$
(VPH).

\begin{figure}
\plotfiddle{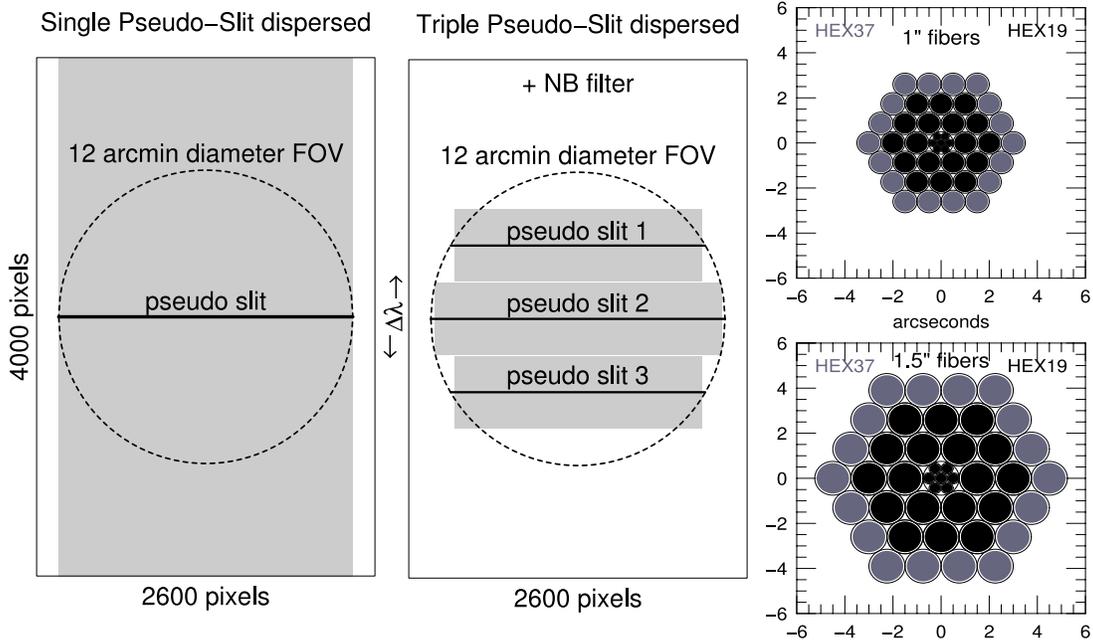}{3in}{0}{85}{85}{-460}{-320}
\caption{The MDX concept for the WIYN Bench Spectrograph and Hydra. At
  left is shown the traditional fiber pseudo-slit with the collimator
  spatial field-of-view (dashed circle), and the camera field-of-view
  (dispersed light; grey) on the 2600x4000 pixel CCD. With the new
  refractive collimator we can use multiple pseudo-slits, each with
  reduced band-pass via a narrow-band pre-filter (middle
  panel). Mini-IFUs of 19-37 fibers (black and grey circles,
  respectively, in right panels) can be positioned in Hydra,
  patrolling 1 deg$^2$. Examples given are for 1'' and 1.5''
  fibers. The total number of mini-IFUs (up to 33) depends primarily
  on the size of the IFUs (\# of fibers) and pseudo-slits.}
\end{figure}

In addition to the Hydra multi-fiber feed, the Bench Spectrograph can
also be fed with single-object fiber integral field-units, including
DensePak (Barden et al. 1998) and SparsePak (Bershady et al. 2004,
2005). These arrays span 30-70 arcsec, sampled at 3-5 arcsec. While
unfortunately the DensePak bundle has recently failed, we plan to
replace it with two new IFUs: HexPak-I \& II, illustrated in Figure
3. They are well suited, for example, to ALFALFA survey follow-up
(Giovanelli, these proceedings). They also offer a
logarithmically-sampled inner core to improve spatial and kinematic
resolution in the centers of galaxies where beam-smearing is
traditionally a problem, dynamical structures are small, AGN can
contribute, and yet there is sufficient flux to enable the finer spatial
sampling.

The IFU concept, however, really achieves its forte for \HI\-survey
follow-up when coupled to the multi-object capacity of traditional
single-fiber positioners. In short, multi-object IFUs covering large
patrol fields can enable high-efficiency follow-up and extension of
the \HI\ aperture-synthesis array surveys noted above. This instrument
idea is not new, e.g., GIRAFFE on VLT (Flores et al. 2004), but the
application needs re-tuning.

Given the depth and area needed to match \HI\ surveys, this instrument
application is best suited to 4m-class telescopes.  As an example,
Hydra could be retrofitted (with available extra slots) to hold 5-12
mini-IFUs spanning 6-9 arcsec in field, sampled at 1-1.5 arcsec per
fiber, and patrolling 1 deg$^2$. The range in number depends on
whether 2 or 3 rings of fibers (19 or 37 total) are desired (i.e.,
field of view) about the logarithmically-sampled core. Two examples
are illustrated in Figure 4 (right-most panels). Such a configuration
would use a single pseudo-slit of fibers, as done with current fiber
feeds.  This delivers a large free spectral range, suitable for
cluster foreground and background studies (field surveys), as well as
abundance work.

Because of the WIYN Bench Spectrograph's new, all-refractive
collimator, however, another exciting possibility emerges. By using
narrow-band filters, the effective band-pass (spectral range) can be
reduced, and multiple pseudo-slits can be introduced. This permits a
trade-off of spectral for spatial multiplex. Consequently this would
allow us to increase the same mini-IFUs up to 14-33 in number, also
patrolling 1 deg$^2$. This so-called MDX concept is outlined in Figure
4. Each fiber would deliver, e.g., $\lambda/\Delta\lambda = 5000$
($\sigma = 25.6$ km s$^{-1}$) with a velocity range of 5600 km
s$^{-1}$, or other commensurate combinations of coverage and
resolution (depending on grating design and angle). This is
well-matched to galaxy cluster studies.

Optical instruments on 4m-class telescopes, such as those described
above, if built, will play a dramatic role in defining the
environmental impact on galaxy evolution at late-times in the
universe. This is a time when clusters are still actively building and
processing galaxies from the blue cloud to the red sequence. The
combined \HI\ and optical views of this transformation -- taken at
redshifts low enough to discern detailed galaxy properties -- offers
an unprecedented opportunity to advance our knowledge of how galaxies
form and evolve.

% Some url test \url{http://www.world.universe}.

\medskip
We acknowledge research support from NSF/AST06-07516. 

%%%%%%%%%%%%%%%%%%%%%%%%%%%%%%%%%%%%%%%%%%%%%%%%
%% The bibliography can be prepared using the BibTeX program or
%% manually.
%%
%% The code below assumes that BibTeX is used.  If the bibliography is
%% produced without BibTeX comment out the following lines and see the
%% aipguide.pdf for further information.
%%
%% For your convenience a manually coded example is appended
%% after the \end{document}
%%%%%%%%%%%%%%%%%%%%%%%%%%%%%%%%%%%%%%%%%%%%%%%%

%%%%%%%%%%%%%%%%%%%%%%%%%%%%%%%%%%%%%%%%%%%%%%%%
%% You may have to change the BibTeX style below, depending on your
%% setup or preferences.
%%
%%
%% For The AIP proceedings layouts use either
%%%%%%%%%%%%%%%%%%%%%%%%%%%%%%%%%%%%%%%%%%%%

% \bibliographystyle{aipproc}   % if natbib is available
%\bibliographystyle{aipprocl} % if natbib is missing

%%%%%%%%%%%%%%%%%%%%%%%%%%%%%%%%%%%%%%%%%%%
%% You probably want to use your own bibtex database here
%%%%%%%%%%%%%%%%%%%%%%%%%%%%%%%%%%%%%%%%%%%
\bibliography{sample}

%%%%%%%%%%%%%%%%%%%%%%%%%%%%%%%%%%%%%%%%%%%
%% Just a reminder that you may have to run bibtex
%% All of it up to \end{document} can be removed
%% if you don't like the warning.
%%%%%%%%%%%%%%%%%%%%%%%%%%%%%%%%%%%%%%%%%%%
% \IfFileExists{\jobname.bbl}{}
%  {\typeout{}
%   \typeout{******************************************}
%   \typeout{** Please run "bibtex \jobname" to optain}
%   \typeout{** the bibliography and then re-run LaTeX}
%   \typeout{** twice to fix the references!}
%   \typeout{******************************************}
%   \typeout{}
%  }

\end{document}